\documentclass{PoS}
\newcommand{\be}{\begin{equation}}
\newcommand{\ee}{\end{equation}}

\title{Decay constants $f_B$ and $f_{B_s}$ from HISQ simulations}

\ShortTitle{Decay constants $f_B$ and $f_{B_s}$}

\author{\vspace{-3mm}
A.~Bazavov$^a$,
C.~Bernard$^b$,
C.~Bouchard$^c$,
N.~Brown$^b$,
\speaker{C.~DeTar} \nolinebreak $^d$,
D.~Du$^e$,
A.X.~El-Khadra$^f$,
E.D.~Freeland$^g$,
E.~G\'amiz$^h$,
Steven~Gottlieb$^i$,
U.M.~Heller$^j$,
J.~Komijani$^b$ \hspace{-1.0mm}\thanks{Present address: Institute for Advanced Study, 
        Technische Universit\"at M\"unchen, Garching, Germany},
A.S.~Kronfeld$^{l,m}$,
J.~Laiho$^e$,
L.~Levkova$^d$,
P.B.~Mackenzie$^l$,
C.~Monahan$^d$ \hspace{-1.0mm}\thanks{Present address: Department of Physics, 
        Rutgers University, New Brunswick, NJ 089001, USA},
T.~Primer$^k$,
Heechang~Na$^d$  \hspace{-1.0mm}\thanks{Present address: Ohio Supercomputer Center, 
        Columbus, OH 43210, USA},
E.T.~Neil$^{n,o}$,
J.N.~Simone$^l$,
R.L.~Sugar$^p$,
D.~Toussaint$^k$,
R.S.~Van~de~Water$^l$, and
R.~Zhou$^l$
\\
\llap{$^a$} Department of Physics and Astronomy, University of Iowa, Iowa City, IA 52240 USA\\
\llap{$^b$} Department of Physics, Washington University, St. Louis, MO 63130, USA\\
\llap{$^c$} Department of Physics, William and Mary University, Williamsburg, VA 23185, USA\\
\llap{$^d$} Department of Physics and Astronomy, University of Utah, Salt Lake City, UT 84112, USA\\
\llap{$^e$} Department of Physics, Syracuse University, Syracuse, NY 13244, USA\\
\llap{$^f$} Department of Physics, University of Illinois, Urbana,  IL 61801, USA\\
\llap{$^g$} Liberal Arts Department, School of the Art Institute of Chicago, Chicago, IL 60603, USA\\
\llap{$^h$} CAFPE and Departamento de F\'{\i}sica Te\'orica y del Cosmos, Universidad de Granada, E-18071 Granada, Spain\\
\llap{$^i$} Department of Physics, Indiana University, Bloomington, IN 47405, USA\\
\llap{$^j$} American Physical Society, One Research Road, Ridge, NY 11961, USA\\
\llap{$^k$} Physics Department, University of Arizona, Tucson, AZ 85721, USA\\
\llap{$^l$} Fermi National Accelerator Laboratory\thanks{Operated by Fermi Research Alliance, LLC, 
under Contract No.~DE-AC02-07CH11359 with the US DOE.}, Batavia, IL 60510 USA\\
\llap{$^m$} Institute for Advanced Study, Technische Universit{\"a}t M{\"u}nchen, 85748 Garching, Germany\\
\llap{$^n$} Department of Physics, University of Colorado, Boulder, CO 80309, USA\\
\llap{$^o$} RIKEN-BNL Research Center, Brookhaven National Laboratory, Upton, NY 11973, USA\\
\llap{$^p$} Department of Physics, University of California, Santa Barbara, CA 93106, USA

\vspace{2mm}
{\large\bf Fermilab Lattice and MILC Collaborations}
\vspace{3mm}

E-mail: 
\email{detar@physics.utah.edu, jkomijani@gmail.com, cb@wustl.edu, doug@physics.arizona.edu}
}

\abstract{We give a progress report on a project aimed at a
  high-precision calculation of the decay constants $f_B$ and
  $f_{B_s}$ from simulations with HISQ heavy and light valence and sea
  quarks.  Calculations are carried out with several heavy
  valence-quark masses on ensembles with 2+1+1 flavors of HISQ sea
  quarks at five lattice spacings and several light sea-quark mass
  ratios $m_{ud}/m_s$, including approximately physical sea-quark
  masses.  This range of parameters provides excellent control of the
  continuum limit and of heavy-quark discretization errors. We present
  a preliminary error budget with projected uncertainties of 2.2~MeV
  and 1.5~MeV for $f_B$ and $f_{B_s}$, respectively.}

\FullConference{The 33rd International Symposium on Lattice Field Theory\\
		14 -18 July 2015\\
		Kobe International Conference Center, Kobe, Japan}

\begin{document}

\begin{figure}
\vspace*{-5mm}
  \begin{centering}
    \begin{tabular}{cc}
      \begin{minipage}{0.475\textwidth}
       \includegraphics[width=0.8\textwidth,clip]{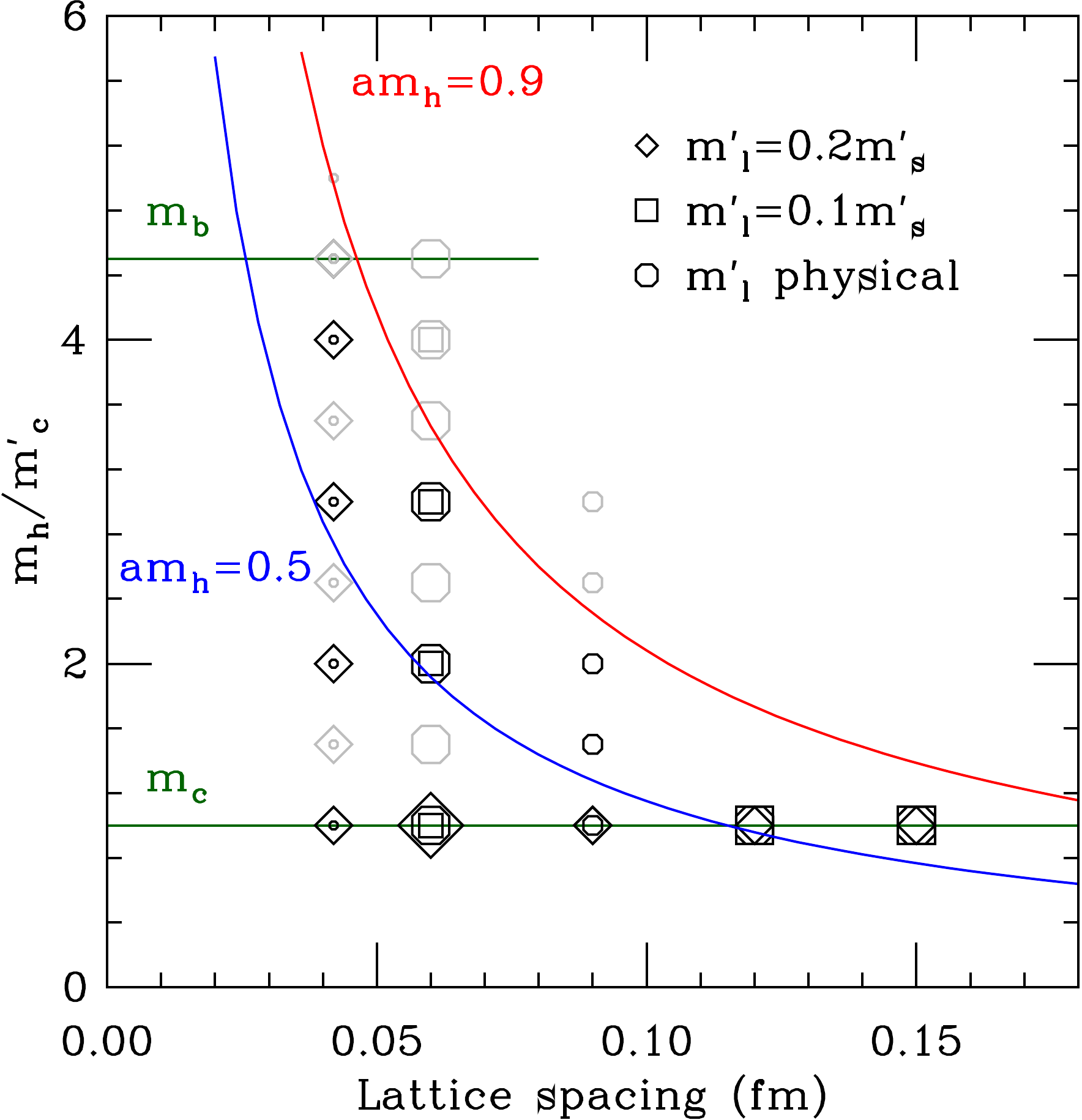}
      \end{minipage}
      &
      \begin{minipage}{0.525\textwidth}
        \includegraphics[width=0.8\textwidth,clip]{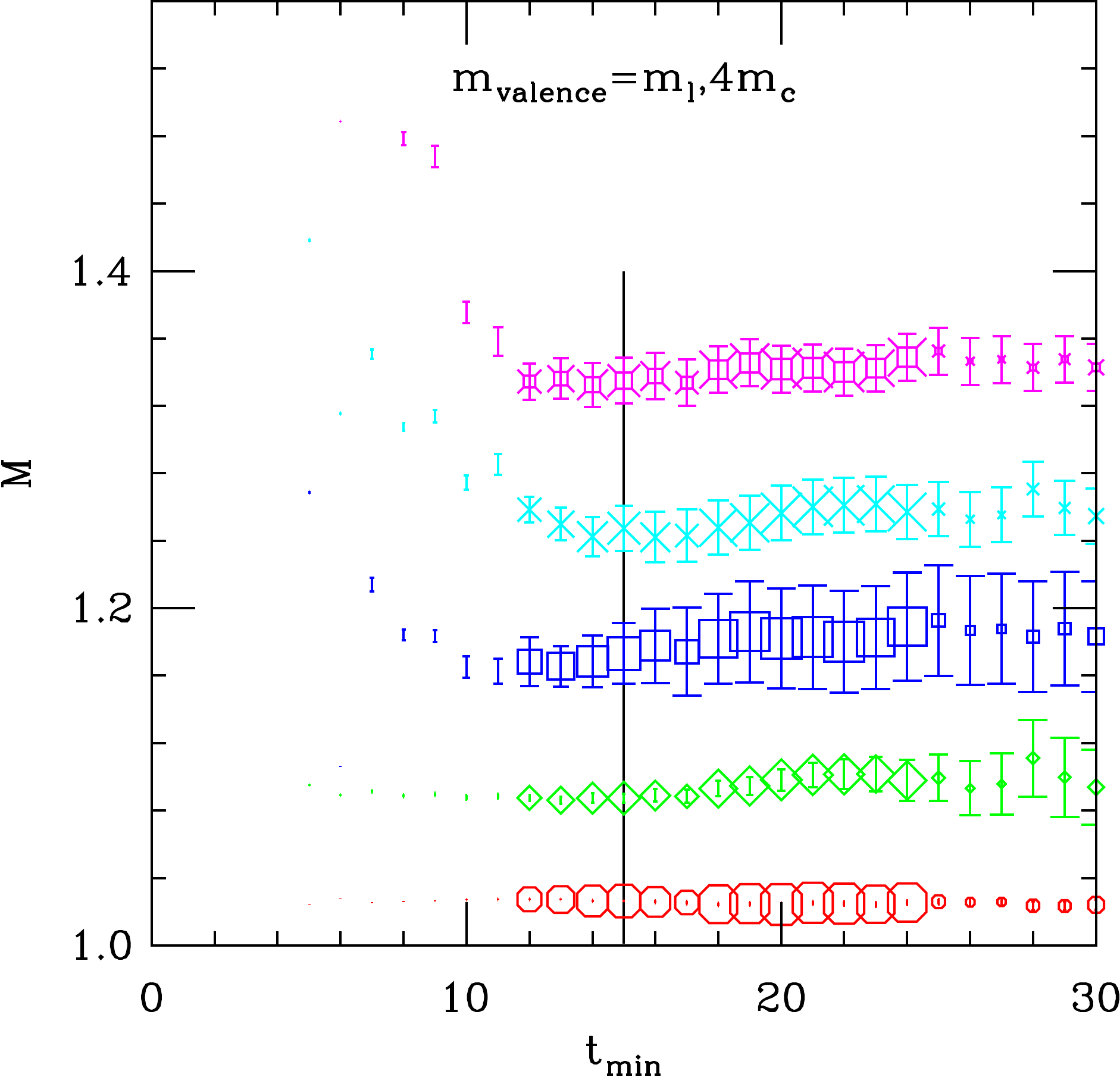}
      \end{minipage}
    \end{tabular}
  \end{centering}
\vspace*{-5mm}
  \caption{
    Left: heavy valence-quark masses and ensemble lattice spacings in
    this study for different light to strange sea-quark-mass ratios
    $m_l^\prime/m_s^\prime$.  The symbol radius is proportional to the
    data sample size.  The red line indicates the cut  $am_h =
    0.9$;  the blue, $am_h = 0.5$.  Faint symbols indicate omitted
    data.
    Right: Spectrum on the 0.042 fm physical-mass ensemble for a
    heavy-light meson with valence masses $4m_c^\prime, m_l^\prime$
    {\it vs.} $t_{\rm min}$, the minimum fit-range distance, with
    $m_l^\prime = m_u^\prime = m_d^\prime$.  The symbol size is
    proportional to the $p$ value.  The inset magnifies the ground
    state.  Priors have been introduced to constrain the excited-state
    contributions.  The vertical line shows our choice for $t_{\rm
      min}$.
 }
  \label{fig:ensembles_spectrum}
\end{figure}

\section{Introduction}

We are searching for new physics by looking for discrepancies between
precise calculations within the Standard Model and high-precision
experimental measurements.  Here we provide a progress report on an
effort to improve Standard-Model calculations of the leptonic decay
constants of the $B$ and $B_s$ mesons, $f_B$ and $f_{B_s}$.  These, in
turn, provide accurate predictions for the charged-current leptonic
decay $B\rightarrow \tau\nu$, for the FCNC decays $B_s \rightarrow
\mu^+\mu^-$, and for neutral $B$ mixing.  Such calculations are also
needed to probe the $V-A$ structure of the $Wub$ vertex and to help
address the present tension between inclusive and exclusive
determinations of the CKM matrix element $|V_{ub}|$.

To achieve high precision, we employ highly improved staggered (HISQ)
quarks \cite{Follana:2006rc,milc_hisq,Bazavov:2010ru} with masses
heavier than the charm quark mass and adopt the HPQCD strategy of
extrapolating to the $b$-quark mass \cite{McNeile:2011ng}.  The
extrapolation is guided by heavy-quark effective theory, which we
incorporate in our chiral/continuum analysis.

We reduce errors from previous calculations for three reasons.  The
staggered-fermion local pseudoscalar density does not require
renormalization, one of the significant sources of systematic error
with other heavy-quark formalisms
\cite{Bazavov:2011aa,Na:2012kp,Dowdall:2013tga,Christ:2014uea,Aoki:2014nga}.
Compared with HPQCD, our ensembles employ HISQ instead of asqtad sea
quarks.  The HISQ ensembles~\cite{Bazavov:2012xda} have a large
statistical sample size, and include lattices with nearly
physical-mass light quarks, which reduces significantly chiral
extrapolation errors.

This work extends our earlier calculation of $f_D$, $f_{D_s}$, which
also used the HISQ formalism for charm quarks \cite{Bazavov:2014wgs}.
Here we include valence fermions heavier than charm. 

\begin{figure}[t]
\vspace*{-7mm}
  \begin{centering}
    \begin{tabular}{cc}
      \begin{minipage}{0.45\textwidth}
       \includegraphics[width=\textwidth,clip]{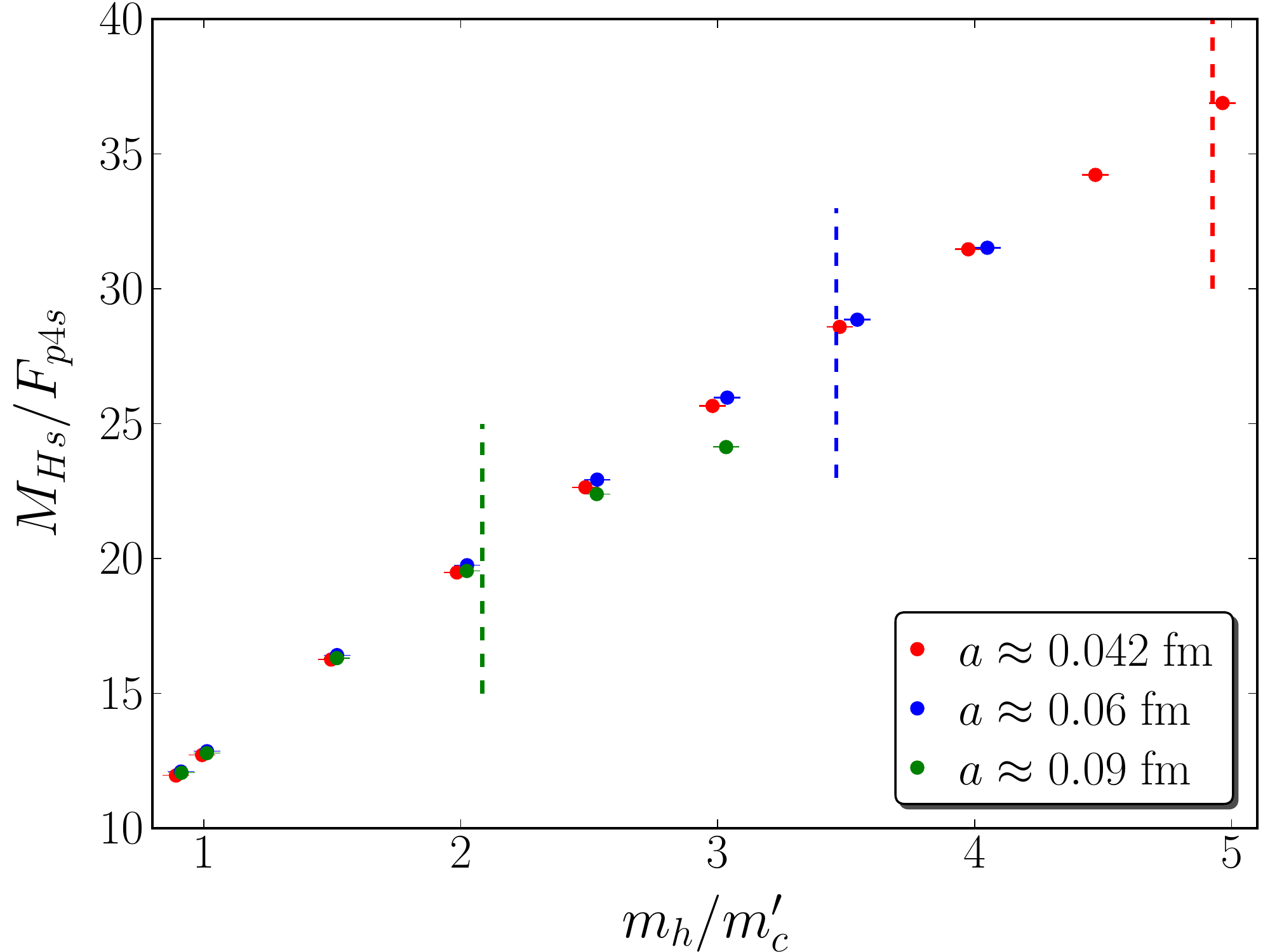}
      \end{minipage}
      &
      \begin{minipage}{0.45\textwidth}
         \includegraphics[width=\textwidth,clip]{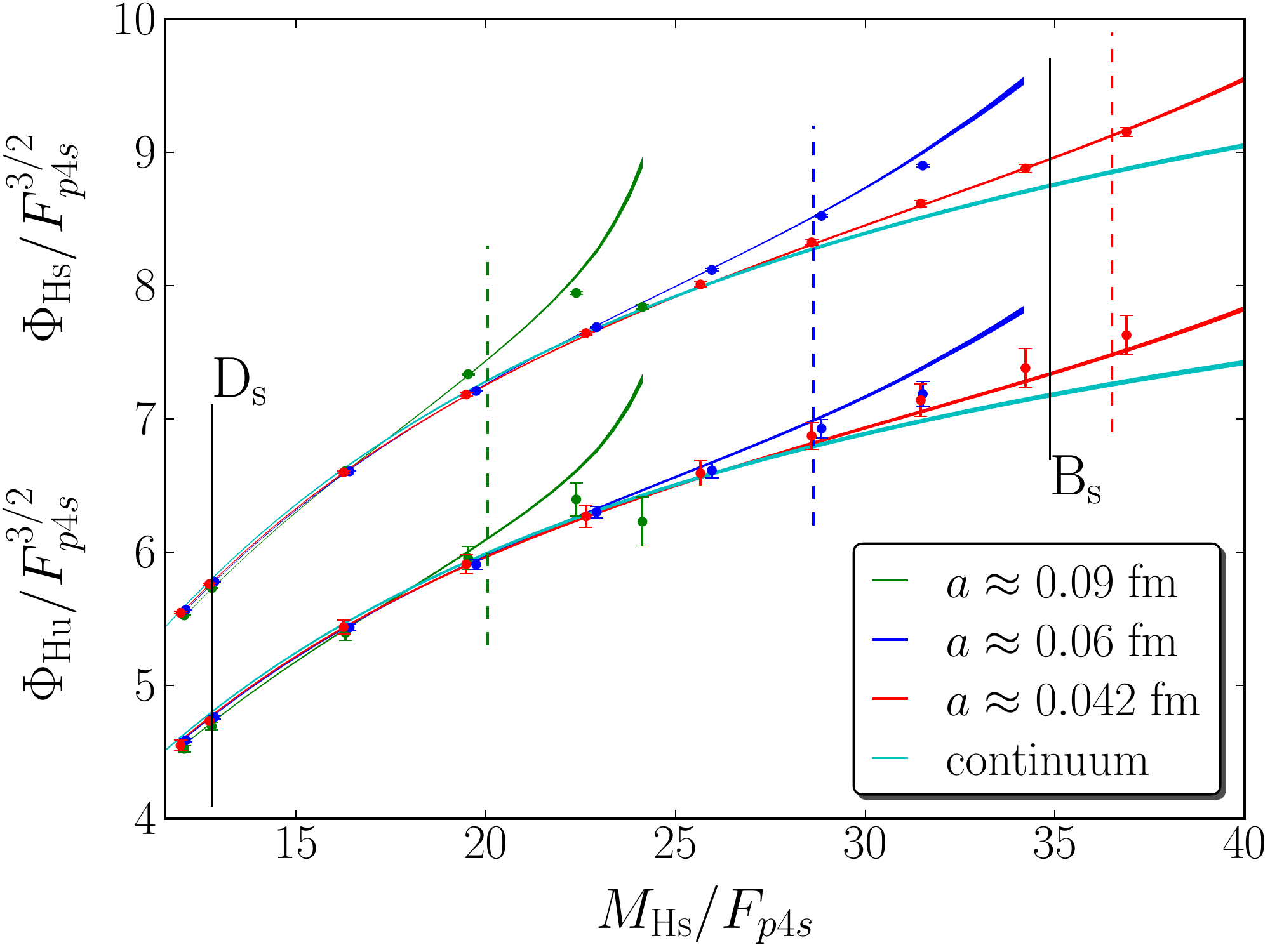}
      \end{minipage}
    \end{tabular}
  \end{centering}
  \caption{ 
   Illustration of heavy-quark discretization effects.
   Left: heavy-strange meson mass in units of $F_{p4s} \approx 154$
    MeV {\it vs.} the ratio of the heavy-quark mass to the simulation
    charm quark mass $m_h/m_c^\prime$ for three lattice spacings at
    physical sea-quark masses. The small differences in the
    heavy-strange mass for different lattice spacings gives an
    indication of the heavy-quark discretization error.
    Right: decay constants $\Phi_{H_q}$ (defined below)
    plotted in units of $F_{p4s}$ {\it vs.} the heavy-strange meson
    mass for three lattice spacings and the continuum extrapolation.
    In both cases the dashed vertical lines indicate the cut 
    $am_h = 0.9$ for each lattice spacing.  Such a pattern of
    deviations was first reported by HPQCD \protect\cite{McNeile:2011ng}.
}
 \label{fig:HQdiscretization}
\end{figure}

The range of valence heavy-quark masses and lattice
spacings for the HISQ ensembles in this study is indicated in the left
panel of Fig.~\ref{fig:ensembles_spectrum}.  HISQ quarks show large
lattice artifacts in the meson dispersion relations when $am_h > 0.9$.
Discretization effects in the heavy-light meson mass and decay
constant, shown in Fig.~\ref{fig:HQdiscretization}, also increase
dramatically for larger $am_h$. So we drop data with $am_h > 0.9$ and
parameterize the heavy-quark mass dependence in our fits at smaller
values with the help of heavy-quark effective theory.  Because of
strong correlations, we also discard some $m_h/m_c^\prime$
points. (Primes on the masses indicate the simulation mass values.)

\begin{figure}[t]
  \begin{centering}
    \begin{tabular}{cc}
      \begin{minipage}{0.5\textwidth}
          \includegraphics[width=0.75\textwidth,clip]{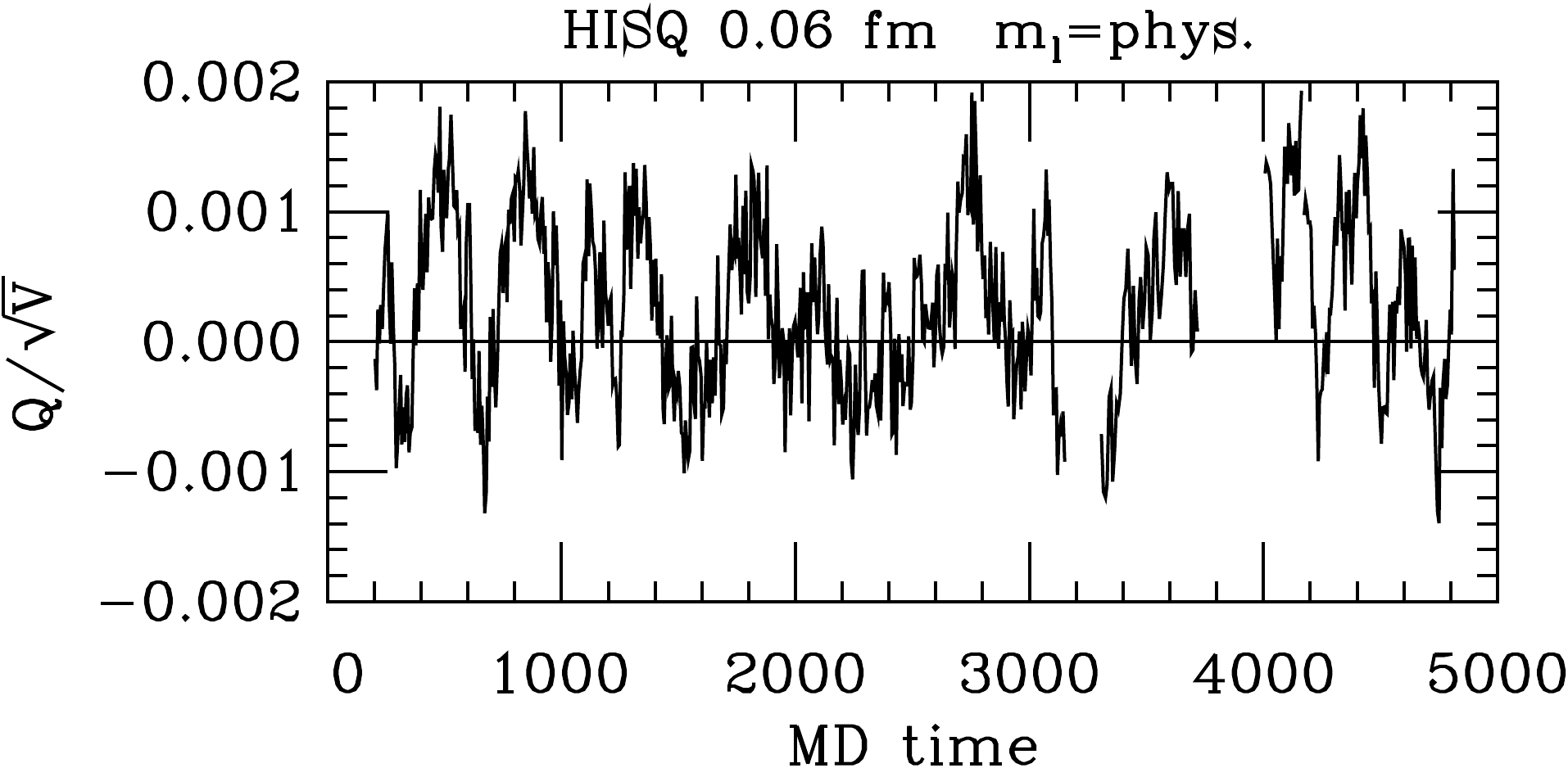}
      \end{minipage}
      &
      \begin{minipage}{0.5\textwidth}
          \includegraphics[width=0.75\textwidth,clip]{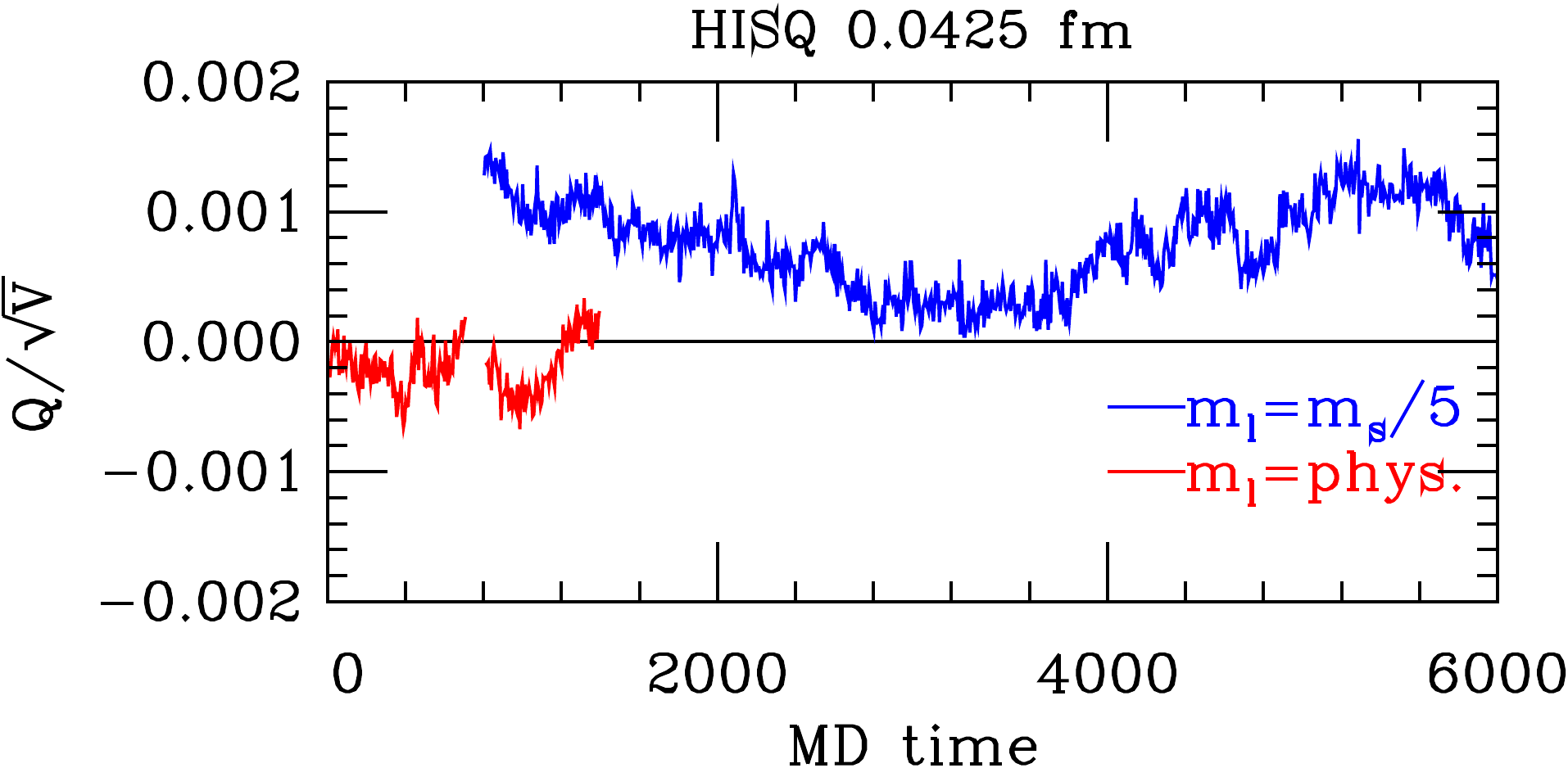}
      \end{minipage}
    \end{tabular}
  \end{centering}
  \caption{Topological charge {\it vs.} molecular dynamics time unit.
    Only fluctuations in $Q^2$ are relevant for the
    (parity-conserving) matrix elements here.
}
\label{fig:topohistory}
\end{figure}

An important question is whether at the smallest lattice spacings our
ensembles are ergodic with respect to the topological charge.  In
Fig.~\ref{fig:topohistory} we show the evolution of topological charge
for our 0.06 fm and 0.042 fm ensembles.  It is clear that tunneling
has slowed at the smaller lattice spacing.  Tunneling is less frequent
when the light sea quark is unphysically heavy. We are currently
investigating the implications for various observables.

\section{Correlator Analysis}

We compute the decay constant from the asymptotic behavior of the
heavy-light pseudoscalar density-density correlator in the usual way:
    \be
      C_{\rm pt-pt}(t) \rightarrow A_{\rm pt-pt} \exp(-M_{H_q} t)
    \ee
where $M_{H_q}$ is the heavy-light meson mass and $q$ labels the light
valence quark.  The decay constant is then obtained without the need
for matching factors from
    \be
      f_{H_q} = (m_h + m_q)\sqrt{\frac{3V A_{\rm pt-pt}}{2M^3_{H_q}}}
    \ee
The correlators are calculated with random and Coulomb-gauge wall
sources and point sinks. We compared fits to 3+2 states (3
nonoscillating and 2 oscillating) and to 2+1 states and conclude that
3+2 states incorporate adequately the effects of excited states. We
can get a sense of the stability of the 3+2 fit from a plot of the
resulting spectrum of ground and excited states as a function of
$t_{\rm min}$, shown in the right panel of
Fig.~\ref{fig:ensembles_spectrum} for the 0.042 fm, physical-mass
ensemble.

\section{Chiral-continuum Extrapolation}

We use heavy-quark effective theory (HQET) to model the heavy-quark
mass dependence of $f_{H_q}$.  We start by relating the QCD current
to the HQET current:

     \be
       J^{\rm QCD} \equiv C_i(m_Q/\mu,\alpha_s(\mu)) J^{\rm HQET}_i(\mu) + {\cal O}(1/m_Q)\, ,
     \ee
where the Wilson coefficient has the perturbative expression
     \be
     C_i(m_Q/\mu, \alpha_s(\mu)) = \left[\frac{\alpha_S(m_Q)}{\alpha_S(\mu)}\right]^{-6/25} 
		      \left[1 + {\cal O}\left(\alpha_S(m_Q), \alpha_S(\mu)\right) \right] \, .
     \ee
So, with $M_{H_s}$ at physical sea-quark masses and physical valence
$s$-quark mass, we remove the scaling factors from the decay constant
and write it in terms of $\Phi_{H_q}$:
\begin{eqnarray}
         f_{H_q}\sqrt{M_{H_q}} &=& \Phi_{H_q} = \left[\alpha_S(M_{H_s})\right]^{-6/25} \left[1 + {\cal O}\left(\alpha_S(M_{H_s})\right) \right] \ \tilde\Phi_{H_q} \, , \nonumber \\
     \tilde\Phi_{H_q} &=& \tilde \Phi_0 \left(1 + k_1\frac{\Lambda_{\rm \small HQET}}{M_{H_s}} + 
            k_2\frac{\Lambda_{\rm \small HQET}^2}{M_{H_s}^2}
      + k_1^\prime \frac{\Lambda_{\rm \small {HQET}}}{m_c^\prime} \right) 
            \left(1 + \mbox{\rm log/analytic terms}\right) \,. \quad
      \label{eq:hqetModel}
\end{eqnarray}
We then introduce the heavy-quark discretization correction for the
low energy constant $\Phi_0$ following HPQCD \cite{McNeile:2011ng}
     \be
      \tilde \Phi_0 \rightarrow \tilde\Phi_0 [ 1 + c_1 \alpha_S(a\Lambda)^2 + c_2 (a\Lambda)^4 + c_3\alpha_S(am_h)^2 + c_4(am_h)^4 + \ldots{}] \, .
     \label{eq:HQdiscCorrect}
     \ee

Finally, for the chiral logarithms and analytic terms in
Eq.~(\ref{eq:hqetModel}), we use heavy-meson rooted all-staggered
$\chi$PT from Bernard and Komijani \cite{Bernard:2013qwa}, which
parameterizes the light-quark mass dependence and incorporates
taste-breaking and generic discretization errors from the light-quark
and gluon actions. Because of space limitations we give only a
schematic representation:
  \begin{eqnarray}
    (1 + \mbox{\rm log/analytic terms}) &=& 1 + \mbox{\rm NLO staggered chiral logarithms} \nonumber \\
                                               &+& \mbox{\rm NLO + NNLO + NNNLO analytic terms}   \label{eq:chiral} \\
    \mbox{\rm NLO analytic terms}  &=& 
     \left(L_s + L_{s,M}\frac{\Lambda_{\rm \small HQET}}{M_{H_s}}\right)\left(2m_l+m_s\right) + \left(L_q + L_{q,M}\frac{\Lambda_{\rm \small HQET}}{M_{H_s}}\right)\left(m_q\right)  + c_{a,\Xi} a^2 \ . \nonumber
  \end{eqnarray}
Note that the coefficients of the NLO analytic terms include a
heavy-quark mass correction.  A heavy-quark mass dependence also
appears implicitly through the $M_{H_q^*} - M_{H_q}$ hyperfine
splitting and heavy-light flavor splittings.

\begin{figure}
  \begin{centering}
    \begin{tabular}{cc}
      \begin{minipage}{0.525\textwidth}
         \includegraphics[width=\textwidth]{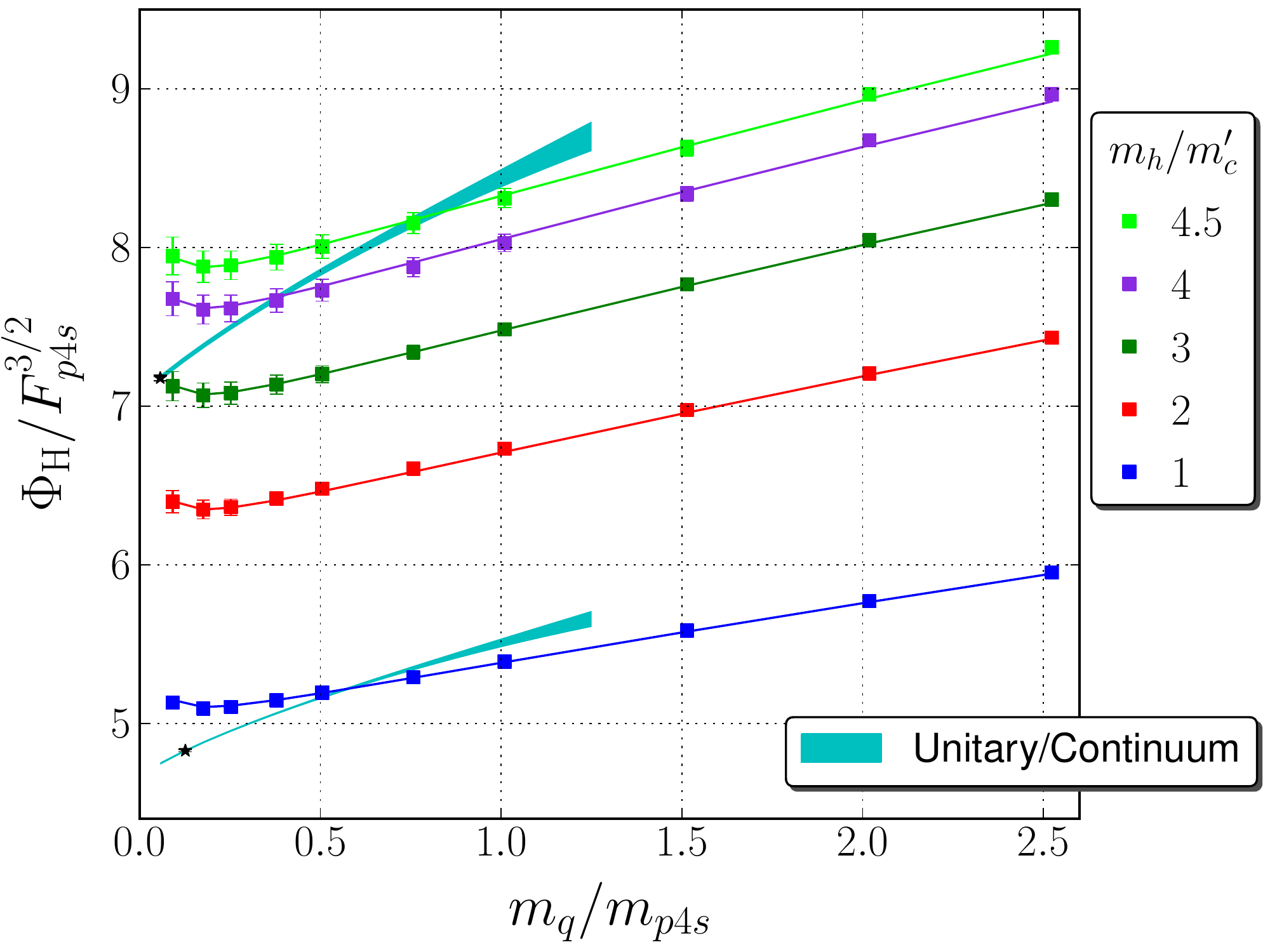} \\
      \end{minipage}
      &
      \begin{minipage}{0.45\textwidth}
         \includegraphics[width=\textwidth]{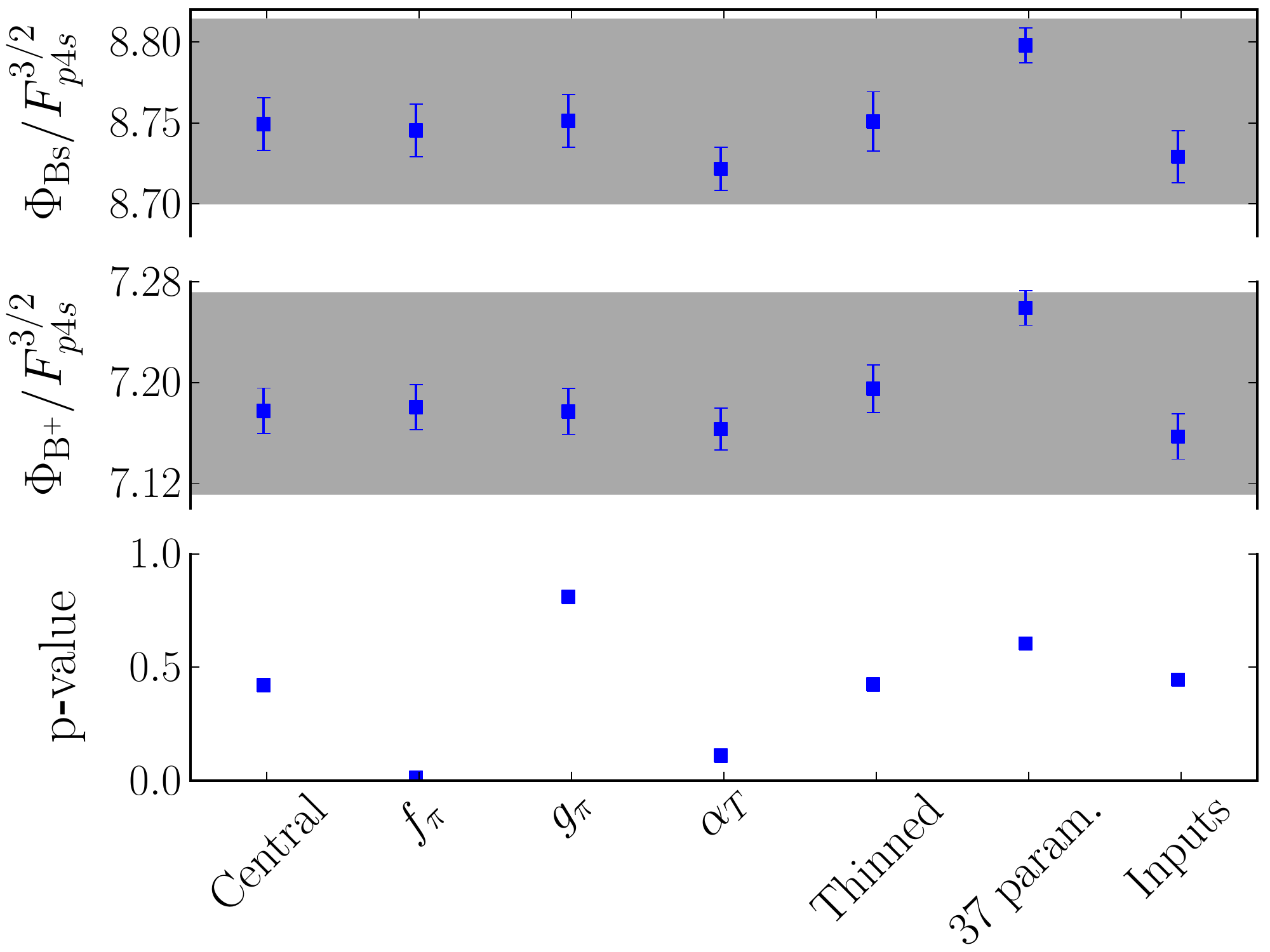} \\
      \end{minipage}
    \end{tabular}
  \end{centering}
\vspace*{-5mm}
  \caption{ Left: sample chiral-continuum fit result for the $a
    \approx 0.042$ fm, $m'_l/m'_s = 0.2$ data.  The unitary/continuum
    result at physical $s$-, $c$-, and $b$-quark masses is shown in
    cyan. The stars indicate the fully physical results for the $B^+$
    and $D^+$ mesons. Right: stability plot showing the sensitivity to
    model choices as described in the text. The error band shows the
    error taken for systematics associated with the chiral-continuum
    fit.  }
  \label{fig:stability}
\end{figure}

Altogether our HQET-chiral continuum-chiral function,
Eq.~(\ref{eq:hqetModel}) with Eqs.~(\ref{eq:HQdiscCorrect}) and
(\ref{eq:chiral}) including NLO, NNLO, and NNNLO terms has 26
parameters.  We use this parameterization for our central fit to our
values of $\Phi_{H_q}$.  An example of the resulting fit is shown in
the left panel of Fig.~\ref{fig:stability}.  From the continuum
extrapolation, we obtain values of the decay constants as a function
of $M_{H_s}$ and the light valence-quark mass.

  \begin{figure}
\vspace*{-5mm}
         \includegraphics[width=0.45\textwidth]{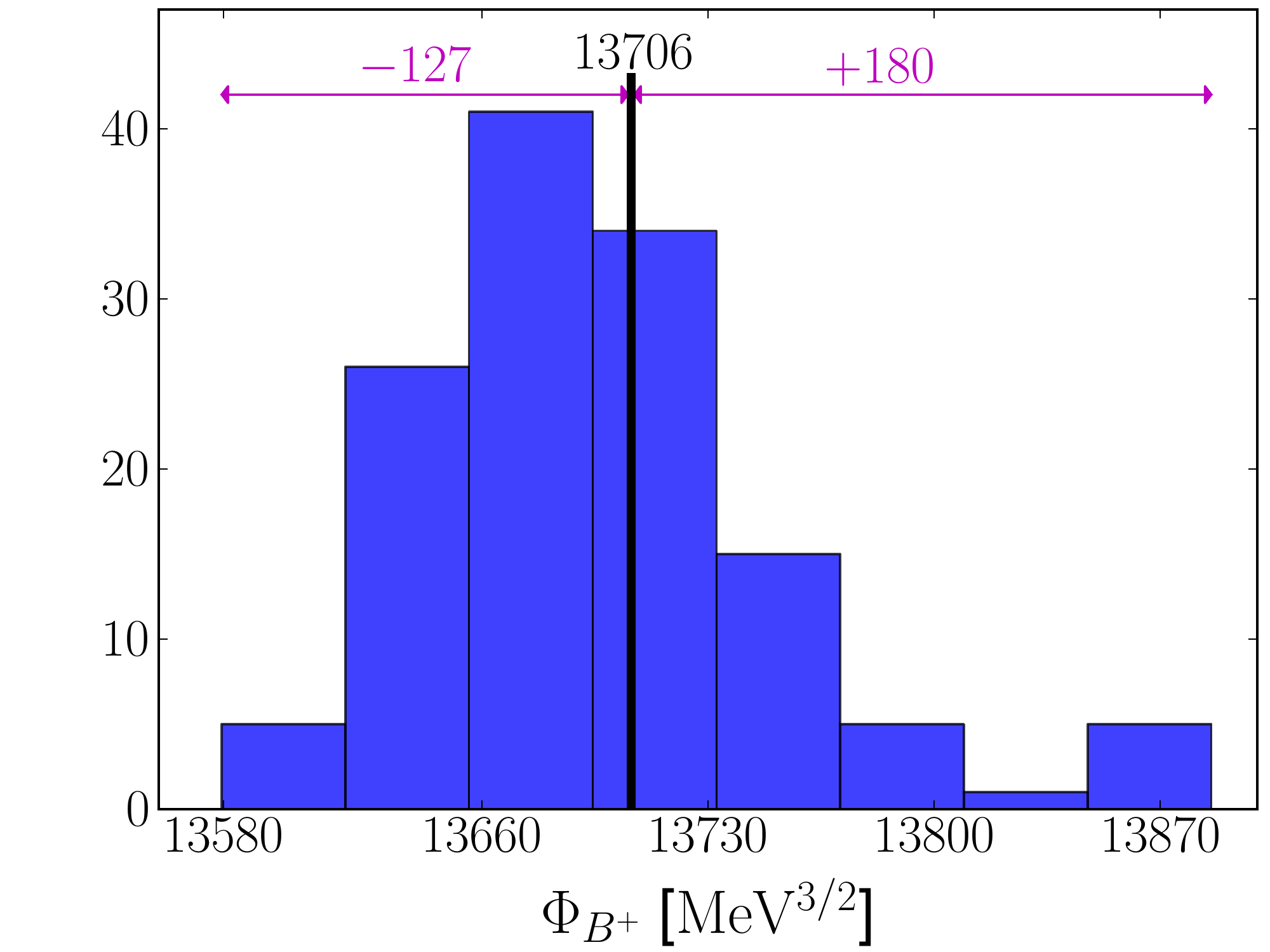} \hfill
         \includegraphics[width=0.45\textwidth]{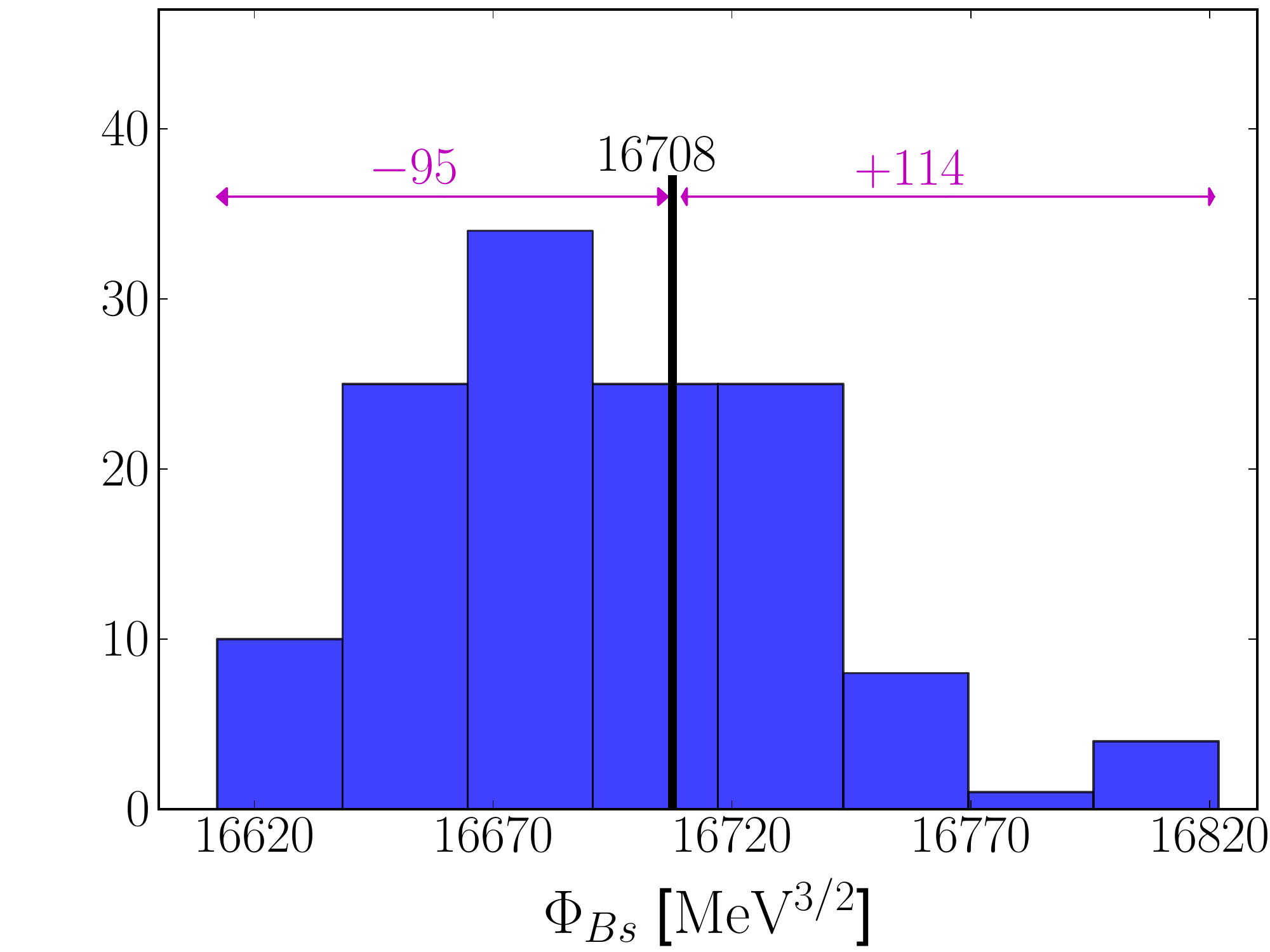}
\vspace*{-3mm}
    \caption{Histogram of central values resulting from varying
      choices of the fit function and input data.  We estimate the
      associated systematic error from the extrema.
}
\label{fig:histo}
  \end{figure}

\section{Error Budget}

To estimate the systematic error in our methodology arising from the
variety of possible analysis choices and to test the stability of our
results, we tried the following alternatives.
      (1) Replaced $f_K$ with $f_\pi$ in the chiral terms;
      (2) Let the $B^*-B-\pi$ coupling $g_\pi$ that enters the 
          chiral logarithms float with a prior 0.53(8).
      (3) Replaced $\alpha_S$ with a value determined empirically from 
          the measured taste splittings.
      (4) Dropped half the light quark masses and refit;
      (5) Increased the number of
        parameters to 37 by adding 11 higher-order heavy-quark
        terms and do not make any SVD cuts in the fit.
      (6) Used an alternative determination of
        the $F_{p4s}$ scale and quark-mass ratios.

The effect of these variations on the decay constant is plotted in the
right panel of Fig.~\ref{fig:stability}.  Selecting 132 combinations
of such variations that yielded $p > 0.05$ gives us the histograms
shown in Fig.~\ref{fig:histo} for $\Phi_B$ and $\Phi_{B_s}$.  We take
the {\em extrema} of the histograms as the systematic error, noting
that the majority of fit variations lie well within this conservative
choice.  The result is shown in the preliminary error budget in
Table~\ref{tab:errbudget}.  The remaining sub-dominant systematic
uncertainties from finite-volume effects, electromagnetic effects, and
the experimental error in $f_\pi$ used to set the absolute lattice
scale are estimated following Ref.~\cite{Bazavov:2014wgs}.  Adding
these errors in quadrature gives the estimated total uncertainty shown
there.
\begin{table}
  \caption{Preliminary error budget.  Note that the first error,
    estimated from the histogram in Fig.~\protect\ref{fig:histo},
    includes the chiral-continuum fit and $g_\pi$ errors, the light-
    and heavy-quark discretization errors, and the error arising from
    excited state contamination.
\label{tab:errbudget}
}
 \begin{center}
  \begin{tabular}{|l|l|l|}
    \hline
     & $f_{B^+}$ (MeV) & $f_{B_s}$ (MeV) \\
    \hline
    $\chi$PT$\oplus g_\pi \oplus$ HQ-LQ disc. $\oplus$ 2-pt fit & 2.1 & 1.4 \\
    Statistics                                                 & 0.6 & 0.5 \\
    Finite volume                                              & 0.3 & 0.2 \\
    Electromagnetic effects                                    & 0.1 & 0.1 \\
    Exp.~$f_\pi$                                               & 0.3 & 0.4 \\ 
    \hline                                                     
    Total                                                      & 2.2 & 1.5 \\
    \hline
    \end{tabular}
  \end{center}
\vspace*{-5mm}
\end{table}

\begin{figure}
\vspace*{-5mm}
         \includegraphics[width=0.5\textwidth]{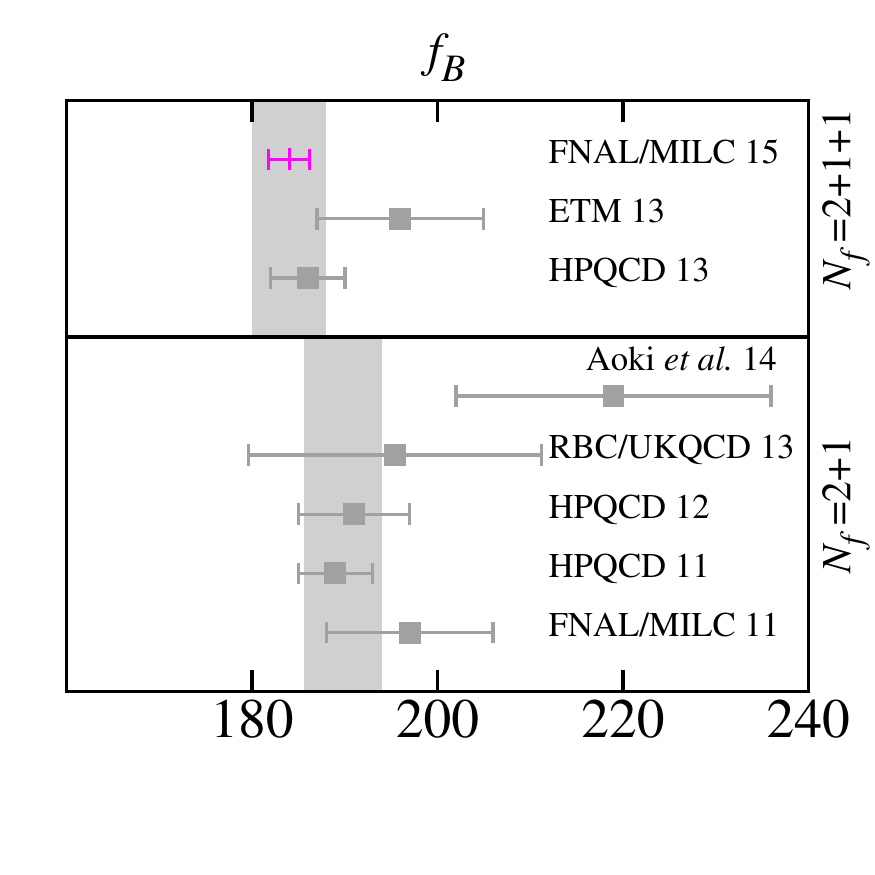}\hfill
         \includegraphics[width=0.5\textwidth]{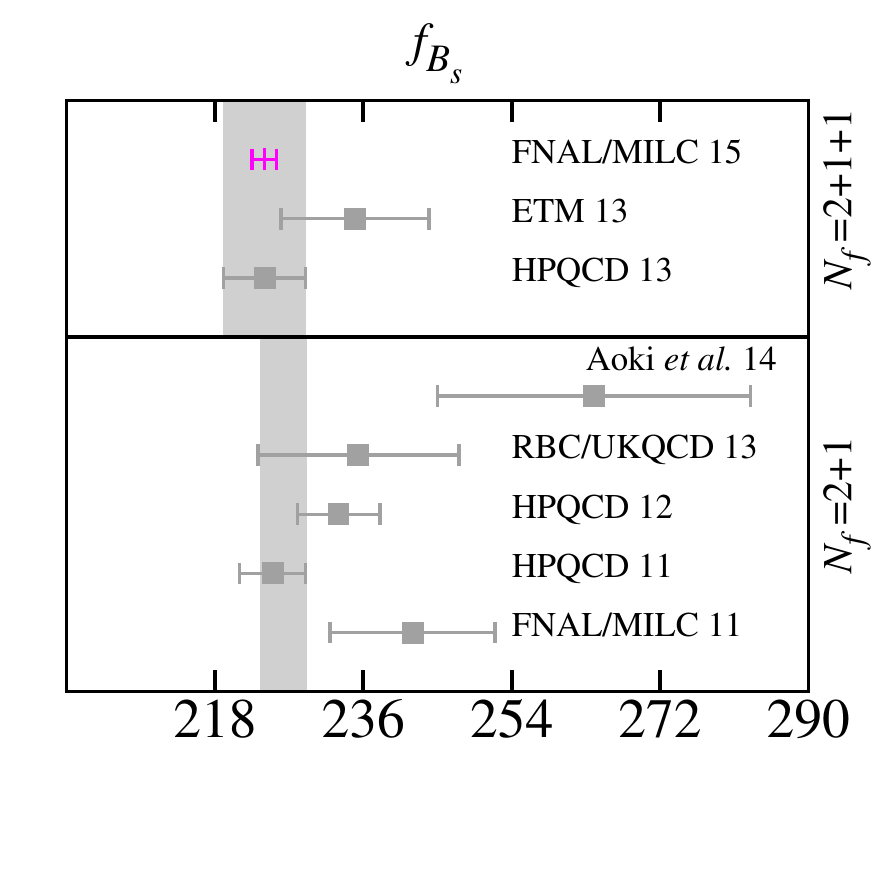}
\vspace*{-10mm}
   \caption{Comparison with other recent
     results~\protect\cite{McNeile:2011ng,Bazavov:2011aa,Na:2012kp,Dowdall:2013tga,Christ:2014uea,Aoki:2014nga,Carrasco:2013naa}.
     Vertical bands show the recent PDG lattice
     averages~\protect\cite{Rosner:2015wva}.  Here we illustrate {\em
       only} our projected error, and take the (2+1+1)-flavor average
     as our central value.  }
 \label{fig:compare}
 \end{figure}

\section{Results and Outlook}
We have presented the status of our analysis of $f_B$ and $f_{B_s}$
from a calculation with all HISQ quarks using the HPQCD scheme of
extrapolating from lighter heavy-quark masses to the $b$-quark mass.
In Fig.~\ref{fig:compare} we compare our projected total error with
the recent compilation of results from the Particle Data
Group~\cite{Rosner:2015wva}.  We anticipate that our calculations when
completed will be the most precise to date. We are presenty generating
and analyzing a 0.03 fm, ($m^\prime_l/m^\prime_s = 0.2$) ensemble for
which $am_b = 0.6$.  Here no extrapolation from lighter heavy-quark
masses is needed.
\acknowledgments

Computations for this work were carried out with resources provided by
the USQCD Collaboration; by the ALCF and NERSC, which are funded by
the U.S. Department of Energy (DOE); and by NCAR, NCSA, NICS, TACC,
and Blue Waters, which are funded through the U.S. National Science
Foundation (NSF).  Authors of this work were also supported in part
through individual grants by the DOE and NSF (U.S); by MICINN and the
Junta de Andaluc\'ia (Spain); by the European Commission; and by the
German Excellence Initiative.


\end{document}